\documentclass[a4paper,twocolumn,aps,showpacs,showkeys,nofootinbib,preprintnumbers,superscriptaddress,amsmath,amssymb,amsfonts]{revtex4}
\usepackage{graphicx}
\usepackage{amsmath}
\usepackage[latin1]{inputenc}
\usepackage[english]{babel}
\usepackage[T1]{fontenc}
\usepackage{amssymb}
\usepackage{amsfonts}
\usepackage{epsfig}
\usepackage{colordvi}
\usepackage{psfrag}
\usepackage{color}
\usepackage{dcolumn}
\usepackage{multirow}
\usepackage{hyperref}
\usepackage{epstopdf}
\usepackage{times}
\hypersetup{
  colorlinks=true,        
  linkcolor=blue,         
  citecolor=magenta,      
}

\newcommand{\aea}{Astron. Astrophys.}
\newcommand{\apjs}{Astrophys. J. Suppl.}

\newcommand{\jcap}{J. Cosmol. Astropart. Phys.}
\newcommand{\mnras}{Mon. Not. R. Astron. Soc.}
\newcommand{\plb}{Phys. Lett. B}

\begin{document}
\title{Tracing the cosmic history by Gauss-Bonnet   gravity}
\author{Ivan de Martino}
\affiliation{Dipartimento di Fisica, Universit\'a di Torino,  Via P. Giuria 1, I-10125 Torino, Italy}
\affiliation{Istituto Nazionale di Fisica Nucleare (INFN), Sezione di Torino, Via P. Giuria 1, I-10125 Torino, Italy}
\affiliation{Donostia International Physics Center (DIPC), 20018 Donostia-San Sebastian (Gipuzkoa) Spain}
\author{Mariafelicia De Laurentis}
\affiliation{Dipartimento di Fisica,  Universit\'a
di Napoli {}``Federico II'', Compl. Univ. di
Monte S. Angelo, Edificio G, Via Cinthia, I-80126, Napoli, Italy}
\affiliation{INFN Sezione  di Napoli, Compl. Univ. di
Monte S. Angelo, Edificio G, Via Cinthia, I-80126, Napoli, Italy.}
\affiliation{Lab.Theor.Cosmology,Tomsk State University of Control Systems and Radioelectronics(TUSUR), 634050 Tomsk, Russia}
\author{Salvatore Capozziello}
\affiliation{Dipartimento di Fisica, Universit\'a
di Napoli {}``Federico II'', Compl. Univ. di
Monte S. Angelo, Edificio G, Via Cinthia, I-80126, Napoli, Italy}
\affiliation{INFN Sezione  di Napoli, Compl. Univ. di
Monte S. Angelo, Edificio G, Via Cinthia, I-80126, Napoli, Italy.}
\affiliation{Gran Sasso Science Institute (INFN), Via F. Crispi 7, I-67100, L' Aquila, Italy.}
\affiliation{Lab.Theor.Cosmology,Tomsk State University of Control Systems and Radioelectronics(TUSUR), 634050 Tomsk, Russia.}
\date{\today}

\begin{abstract}
Cosmic history can be traced  considering further curvature contributions inside the gravitational action. Assuming that standard General Relativity can be extended by other curvature invariants, we discuss the possibility that an action containing higher-order curvature terms can fit, in principle, the whole universe evolution. In particular, a theory like $F(R,{\cal G})$, with $R$ the Ricci scalar and $\cal G$ the Gauss-Bonnet topological term, contains  all the curvature invariants that, depending on the energy regime, can address inflation, matter dominated and dark energy regimes. In this paper, we investigate this possibility considering how $F(R,{\cal G})$ models can lead gravity from ultraviolet to  infrared scales. Specifically, we will take into account a cosmographic approach for this purpose.

 \end{abstract}
 \pacs{98.80.-k, 95.35.+d, 95.36.+x}
\keywords{Alternative theories of gravity;  dark energy; observational data; cosmography.}
 
\maketitle
\section{Introduction }\label{uno}
According to our current perspective, the Universe seems dominated by two ingredients we find frustratingly difficult to understand at fundamental level. One of these is the {\it dark matter} which drives the formation of self-gravitating systems. Despite extensive searches for new candidates beyond the Standard Model of Particle Physics that could account for the discrepancy between luminous and non-luminous matter, no final evidence   emerged so far \cite{Tanabashi:2018oca}. 

Furthermore, the  recent issue of the accelerated expansion of the Hubble flow gave rise to the possibility of another elusive component of cosmic pie, the so-called {\it dark energy}. Also in this case, no fundamental particle, capable of addressing the cosmological dynamics, has been detected  so far (for a comprehensive review see, for instance, Refs. \cite{Brax:2017idh,2018RPPh...81a6901H}). On one hand, we need a mechanism able to cluster structures and, on the other hand, we need a mechanism able to speed up the cosmic fluid. Despite the existence of macroscopic evidences, no fundamental ingredient emerged till now to account for the  $95\%$  of cosmic matter-energy content (for a comprehensive review we refer to \cite{2020arXiv200108297G}). 

This state of art triggered the development of alternative theories of gravity formulated with the aim to replace, correct or extend General Relativity (GR) in view of accounting for the discrepancies in apparent mass and cosmic acceleration. Nevertheless GR  explains a broad range of phenomena  and it works very well to describe the Universe as a whole, provided dark matter and dark energy exist as separate entities. According to this fact, any alternative theory has to reproduce GR results trying to  account for additional effects as those detected at infrared scales \cite{2015Univ....1..123D, Annalen,2018PhRvD..97j4067D, 2018PhRvD..97j4068D, 2018EPJC...78..916D}. Furthermore, the lack of a self-consistent quantum gravity  drives the search for effective theories \cite{Hogan} as  alternatives
 at ultraviolet scales. These approaches led to a {\it semi-classical} picture where geometry is described by a space-time continuum and matter-energy side is given by some scalar fields or geometric corrections \cite{Buchbinder1992,Birrell}. However, the picture is consistent if we are quite far from the Planck scales and matter-energy can be averaged to give classical counterparts. 

 One of the common features of these effective models is including invariant terms and scalar fields in the gravitational action \cite{Rept}. In particular, curvature invariants, constructed by Ricci, Riemann and Weyl tensors, emerge as soon as renormalization and regularization of quantum fields on curved spaces are considered \cite{Buchbinder1992, Birrell}. As it is well-known, these terms have a cosmological impact and can trigger inflation at early epochs \cite{Starobinsky1980} and dark energy at late epochs \cite{Quintessence}.

The basic idea is that gravity agrees with GR  at certain scales and epochs   but may have  quite different behaviors at very short (ultraviolet) and very large (infrared) scales. In principle, considering corrections related only with scalar  curvature, one can expect  effective actions like \cite{Sotiriou}
\begin{equation}
F(R)=..+\frac{\alpha_2}{R^2}+\frac{\alpha_1}{R}+\alpha_0+\frac{R}{\beta_1}+\frac{R^2}{\beta_2}+\frac{R^3}{\beta_3}+...,
\end{equation}
where negative powers work at infrared  while positive powers work at ultraviolet scales \cite{Power}. Here $\alpha_0$ has the role of a cosmological constant, and $\alpha_i$ and $\beta_i$ are dimensional couplings. However, this is a phenomenological approach  giving just a coarse-grained picture.  In a more refined theory other curvature invariants should be considered. 
 
According to the previous considerations, issues as the trace anomaly or the renormalization at one-loop level require higher-order curvature invariants which are mostly related to the recovery of effective gravitational actions in curved space time \cite{Buchbinder1992, Birrell}. Specifically, these corrections to the Hilbert-Einstein action are second-order curvature invariants giving rise to fourth-order field equations in metric derivatives. Considering a generic theory containing curvature invariants means to take into account Lagrangians like $F(R,R_{\mu\nu}, R_{\mu\nu\delta\gamma})$ and then improving the number of degrees of freedom related to the gravitational field. However, as discussed in details in \cite{Bogdanos,Francesco,Dialektopoulos}, surface terms coming from combinations of curvature invariants can reduce the complexity of these theories.

A specific role is played by the Gauss-Bonnet topological term $\cal G$ which is a second order combination of curvature invariants defined as
\begin{equation}
{\cal G}\equiv
R^2-4R_{\alpha\beta}R^{\alpha\beta}+R_{\alpha\beta\rho\sigma}
R^{\alpha\beta\rho\sigma}\,,
   \label{GBinvariant}
\end{equation}
where $R$ is the curvature scalar, $R_{\alpha\beta}$ is the Ricci tensor and $R_{\alpha\beta\rho\sigma}$ is the Riemann tensor. In differential geometry, it is  
\begin{equation}
\int_{\cal{M}} {\cal G} d^nx = \chi(\cal{M})\,,
\end{equation}
where $\chi(\cal{M})$ is the Euler characteristic of a manifold $\cal M$ in $n$ dimensions. For $n=4$, $\chi({\cal M})=0$   so it can be considered a surface term not affecting dynamics. 
Furthermore, $\cal G$ emerges in the trace anomaly as soon as one wants to regularize and renormalize gravity at one-loop level (see \cite{Rept,Birrell} for details). However,  for any non-linear function 
of ${\cal G}$, this property does not hold and then contributions to the gravitational action are non-trivial \cite{Nojiri2005}. 

According to these considerations, a generic function $F(R,{\cal G})$ can contain the whole information related to fourth-order dynamics. In other words, a two-scalar field theory with combinations of  $R$ and ${\cal G}$  well represents gravity with second-order curvature invariants.

This class of models is capable of describing several phenomena at different scales, such as, the current acceleration of the Universe  at late epochs and  double inflation at early epochs \cite{DeLaurentis2015a, DeLaurentis2015b}. Strong field phenomena, like  extreme neutron star configurations, can be also framed in the context of Gauss-Bonnet gravity  \cite{Astashenok}.
Moreover, this kind of theories can satisfy the Solar System tests in the weak field limit \cite{DeLaurentis2014}.

In general, Gauss-Bonnet gravity can be relevant  in cosmology as shown in 
\cite{Nojiri2005,Nojiri2007,Nojiri2006,Bazeia2007,Cognola2007,Barrow2007,Bazeia2008,Goheer2009a,Goheer2009b,Mohsenir2010,Sadjadi2010,Alimohammadi2009,Boehmer2009,Uddin2009,Zhou2009,Ivanov2012,Nojiri2010,KumarSanyal2011,Grandal2012,Grandal2013,Benetti1,Benetti2}.

In this perspective, some important issues can be the following: being such a theory well motivated at UV and IR scales, is it possible to select Gauss-Bonnet gravity models capable of tracing  the Universe history at any epoch? In some sense, is it possible  to connect inflation and dark energy epochs, passing through matter epoch, without choosing "ad hoc" models \cite{Odi1,Vasilis}? 

These questions may find an answer using the cosmographic approach \cite{Weinberg1972,Visser2005} that can be extended also to higher redshift regimes \cite{cattoen2007hubble,Orlando,Benetti3} and, in principle, is able to constrain the value of the Hubble constant and the other cosmographic parameters. 
Once these parameters are determined, they can be used to fix reliable constraints on theoretical models. 
This method is especially suited to the study of higher order gravity theories \cite{2008PhRvD..78f3504C,Orlando2} like the one we shall study in the next sections.

Obviously, considering higher order theories  with more than one field, such as $F(R,{\cal G})$, involves some mathematical difficulties in handling field equations and therefore also in finding analytical expressions for the cosmographic parameters.
The choice of the function $F(R,{\cal G})$ is crucial in this type of approach. Therefore, assuming very general hypotheses like analyticity and derivability, one is able to obtain useful relations among cosmographic parameters and the $n$-th derivatives $F^{(n)}(R,{\cal G}) = d^nF/d(R^n{\cal G}^n)$ for any  choice of the function $F (R,{\cal G})$.

The paper is organized as follows. In Sec. \ref{due}, we sketch $F(R,{\cal G})$ gravity deriving its field equations. Sec.\ref{tre} is devoted to Gauss-Bonnet cosmology. In particular we derive the Friedman equations that will be used in the following analysis. In Sec.\ref{quattro}, we discuss how $\cal G$ terms are particularly useful in early epochs and, together with $R^2$ corrections, give rise to inflation. The cosmographic approach is developed in Sec. \ref{cinque} giving all the technical details of the cosmographic series. Observational data and the methodology adopted for the analysis are described in Sec. \ref{sei}.  
In Sec. \ref{sette}, the $F(R,{\cal G})$ cosmography is derived together with the  observational constraints on the models. Discussion and conclusions are drawn in Sec. \ref{Conclusions}. In the Appendix \ref{app:A}, calculations adopted for the cosmographic analysis are reported.

\section{ Gauss-Bonnet Gravity} \label{due}

Let us start by writing a general action for Gauss-Bonnet gravity \footnote{We are using physical units.} \cite{DeLaurentis2014,DeLaurentis2015a,FelixOdin}
\begin{equation}
{\cal A}=\frac{1}{2\kappa}\int d^4x \sqrt{-g}F(R,{\cal G})+{\cal A}_M\,,
   \label{action}
\end{equation}
where $\kappa=8\pi G$, ${\cal A}_M$ is the standard matter action, and $F(R,{\cal G})$ is a function of the Ricci scalar and the Gauss-Bonnet topological invariant.
Variation of the action (\ref{action}) with respect to the metric provides the following gravitational field equations
\begin{equation}\label{eom}
R_{\mu \nu}-\tfrac{1}{2} g_{\mu \nu}R=\kappa\,T^{(mat)}_{\mu \nu}+T^{(\cal{GB})}_{\mu \nu},
\end{equation}
where $T^{(\cal{GB})}_{\mu \nu}$ is defined as
\begin{align}
T^{(\cal{GB})}_{\mu \nu} & =\nabla_\mu \nabla_\nu  F_R-g_{\mu \nu}\, \Box\,  F_R+2R \nabla_\mu \nabla_\nu  F_{\cal G} \nonumber\\
& -2g_{\mu \nu} R\, \Box\,  F_{\cal G}-4R_\mu^{~\lambda} \nabla_\lambda \nabla_\nu  F_{\cal G}-4R_\nu^{~\lambda} \nabla_\lambda \nabla_\mu  F_{\cal G}\nonumber\\
&+4R_{\mu \nu} \Box  F_{\cal G}+4 g_{\mu \nu} R^{\alpha \beta} \nabla_\alpha \nabla_\beta  F_{\cal G}\nonumber\\
&+4R_{\mu \alpha \beta \nu} \nabla^\alpha \nabla^\beta  F_{\cal G}-\frac{1}{2}\,g_{\mu \nu} \left( R F_R + {\cal G}  F_{\cal G}- F\right) \nonumber\\
&+\left(1- F_R\right)\,\left(R_{\mu \nu}-\frac{1}{2} g_{\mu \nu}R\right)\,,
\label{effective-energy-momentum}
\end{align}
involving all extra terms with respect to GR.
As shown in \cite{Mantica}, $T^{(\cal{GB})}$ can be recast as a perfect fluid of geometric origin.

The trace equation is 
\begin{eqnarray}\label{trace}
&&- 2 F + F_R R + 3 \nabla^2 F_R + 2 F_{\cal G}{\cal G} + 2 R \nabla^2 F_{\cal G}\nonumber\\&& - 4 R_{\rho \sigma}
\nabla^\rho \nabla^\sigma F_{\cal G} = 2 \kappa^2 T.
\end{eqnarray}
and derivatives are denoted as 
\begin{equation}
  \label{eq:def1}
  F_{R}\equiv\frac{\partial F(R,{\cal G})}{\partial R}\,,\qquad F_{\cal G}\equiv \frac{\partial F(R,{\cal G})}{\partial {\cal G}}.
\end{equation}
GR is recovered as soon as $F(R,{\cal G})\rightarrow R$. If ${\cal G}$ is not taken into account,  we recover immediately $F(R)$ gravity. In this sense, Gauss-Bonnet cosmology is a straightforward two-field extension  of $F(R)$ gravity. 

\section{Gauss-Bonnet Cosmology}
\label{tre}

Starting from the above theory, it is possible to derive the related cosmology.  We
 consider a  spatially flat Friedman-Robertson-Walker  (FRW) metric like
\begin{equation}
ds^{2}=-dt^{2}+a^{2}(t)(d{x}^{2}+d{y}^{2}+d{z}^{2})\,,
\label{metric}
\end{equation}
where $a(t)$ is the scale factor of the Universe. In this background,  the cosmological equations (see for details Eqs. \eqref{eq:Hsqr}-\eqref{eq:Hdot}), without standard matter,  can be written in a simplified way as \cite{FelixOdin,DeLaurentis2015b}
\begin{eqnarray}\label{FRWH}
&&\dot{H}= \Psi \left[H \dot{F}_R-\ddot{F}_R+4H^3\dot{F}_{\cal G}-4H^2\ddot{F}_{\cal G}\right],
\\
\label{energy}
&&H^2=
\frac{\Psi}{3}\left[F_RR-F(R,{\cal G})-6H\dot{F}_R+{\cal G}F_{\cal G}\right],
\end{eqnarray}
where we have defined
\begin{equation}
    \Psi \equiv \frac{1}{2 F_R + 8H \dot{F}_{\cal G}}\,.
\end{equation}
Furthermore two Lagrange multipliers, defining $R$ and $\cal G$ as functions of $a$ and its derivative $\dot a$ and $\ddot a$, have to be considered to complete the dynamical system. See \eqref{laglag} 
reported in  Appendix.

This system of equations  will be the starting point to develop our considerations at early epoch. As said above, our aim is to track back the cosmic history investigating the curvature regimes related to $F(R, {\cal G})$ gravity. As we shall see, combining information coming from the behavior of $\cal G$ and $R$ functions can, in principle, 
give a self-consistent picture of cosmic evolution at any epoch.

\section{Early epoch cosmology}
\label{quattro}
A first consideration is related to the fact that, as pointed out by Starobinsky \cite{Starobinsky1980}, adding higher-order curvature invariants in the effective action gives rise to inflationary episodes that naturally emerge  without introducing {\it ad hoc} inflaton fields. In other words, improving geometry by curvature invariants give  the possibility to obtain accelerated expansions. 

In the present case, the Ricci and Gauss-Bonnet invariants play the role of scalar fields, whose dynamics is determined by a Klein-Gordon-like equation coming from the trace Eq. \eqref{trace}. In fact, such an equation can be recast as \cite{DeLaurentis2015a}
\begin{equation}
\label{klein}
3\left[\Box F_R+V_R\right]+R\left[\Box F_{\cal G} +W_{\cal G}\right]=0\,,
\end{equation}
according to \eqref{eq:def1} and the definition
\begin{equation}
\label{pot}
V_R=\frac{\partial V}{\partial R}=\frac{1}{3}[R F_R-2F(R,{\cal G})]\,,\quad
W_{\cal G}=\frac{\partial W}{\partial {\cal G}}=2\frac{\cal G}{R}F_{\cal G}\,.
\end{equation}
Here $\Box$ is the d'Alembert operator in curved space-time. Clearly, as demonstrated in \cite{DeLaurentis2015a}, the dominance of one of the two terms in square brackets determines the  evolution related to $R$ or $\cal G$.
Consequently, we expect a double  inflation in which both geometric fields play a role at different scales. In other words, we can have a $R$-dominated inflation and a further $\cal G$-dominated inflation working at different scales. 

{However, before starting our discussion,   the possible presence
of ghosts is an issue  to be considered  in the framework of  Gauss-Bonnet gravity. It is well known that, in general, higher-order theories of gravity are not ghost-free so that  viable ranges of parameters have 
to be selected in order to obtain self-consistent models. This is particularly relevant in cosmology  to obtain reliable cosmic histories to be matched with data. Specifically, higher-derivative gravity contains ghosts  due to the Ostrogradsky instability \cite{Woodard, Suyama}.} 

{These  ghost terms may occur at fundamental and  cosmological  level for $F(R, {\cal G})$ gravity and can be parameterized by  superluminal modes $k^4$  where $k$ is the  wavenumber of the specific mode.  The reason why one gets $k^4$ is due to the fact that $R$ evolves as  $\ell ^{-2}$ and ${\cal G}$ as $\ell ^{-4}$.}

{To obtain viable models,  procedures to  eliminate ghosts have  been developed. In \cite{VasilisGB}, it is proposed a method based on the introduction of an auxiliary scalar field  $\chi = \chi ({\cal G})$ into the $F(R,{\cal G})$ action. Such a field quantifies the propagation of scalar modes. Then, in order to make the  scalar mode not  a ghost,
a canonical kinetic term of $\chi$ can be introduced in the $F(R, {\cal G})$ action. See also   \cite{Sasaki}.  According to this procedure, it is possible to obtain second order field equations, (and then eliminate  higher than second order derivatives) and  impose suitable  initial conditions   determining  a regular and unique  evolution without  ghost fields.
In particular, the  new dynamical degree of freedom can be eliminated adopting a "mimetic gravity" procedure by introducing a Lagrange multiplier \cite{mimetic1,mimetic2,mimetic3}. Such a multiplier gives a natural mass constraint by which the kinetic term becomes a constant.  This constraint determines the range of parameters where the model is ghost-free. Clearly, being the Gauss-Bonnet terms  of fourth-order, the mass parameter has to be $\mu^4$. 
See \cite{VasilisGB} for details.}

With the above considerations in mind, we can take into account  the simplest natural extension of the Starobinsky model \cite{Starobinsky1980} adding a non-trivial  of Gauss-Bonnet contribution, that is \cite{DeLaurentis2015a}:
\begin{equation}\label{starotopo}
{F(R,{\cal G})}= R+\alpha R^2+\beta {\cal G}^2\,,
\end{equation}
where $\alpha$ and $\beta$ are coupling constants with dimensions $\ell^2$ and $\ell^4$, respectively.
This choice means that Gauss-Bonnet invariant gives a further scale where curvature can play a relevant role.
In a scenario where the Universe is homogeneous and isotropic, it is ${\cal G}^2 \sim R^4$ and this allows us to rewrite the above  function  as 
\begin{equation}
F(R) \simeq R + \alpha R^2 + \beta R^4\,,
\end{equation}
which is an improved $F(R)$ model with respect to the quadratic Starobinsky one. 
 Obviously,  considering anisotropies and inhomogeneities, $ {\cal G} ^ 2 \neq R ^ 4 $ due to the fact that we cannot neglect the extra diagonal components of the Ricci and Riemann tensors.
 
In general, an inflationary behavior is achieved if the following
conditions on the Hubble parameter and its derivatives are  satisfied: 
\begin{equation}
\left |\frac{\dot{H}}{H^2}\right|\ll1\,,  \hspace{1cm} \left |\frac{\ddot{H}}{H \; \dot{H}}\right|\ll1\,.
\end{equation}
From the energy condition, given by Eq. 
\eqref{energy}, we have
\begin{eqnarray}\label{Hquad}
&&12 \alpha H \ddot{H}+ H^2+36 \alpha H^2 \dot{H}+ 288 \beta H^4 \dot{H}^2  \nonumber \\
&&+192 \beta H^5 \ddot{H} + 576 \beta H^6 \dot{H}-96 \beta H^8 - 6 \alpha \dot{H}^2=0\,,
\end{eqnarray}
and  from  \eqref{FRWH}, we obtain
\begin{eqnarray}\label{Hdot}
&& 576\beta H^2{\dot H}^3+768\beta H^3{\dot H}{\ddot H}+\beta H^4\left(1728{\dot H}^2+96{\dddot H}\right)
\nonumber\\&&+288\beta H^5{\ddot H}-384\beta H^6 {\dot H}^2
\nonumber\\&&+18\alpha H {\ddot H}+24 \alpha {\dot H}^2+6\alpha {\dddot H}+{\dot H}
=0\,.
\end{eqnarray}
Imposing that $H$ is slowly varying, which means  ${\dot H}<<H^2$ and ${\ddot H}<<H{\dot H}$,  Eq.\eqref{Hquad}, takes the form
\begin{eqnarray} \label{Hquad1}
&& H^2+6\alpha\left(2H{\ddot H}+6H^2{\dot H}-{\dot H}^2\right)\nonumber\\&&
+96\beta H^4\left(3{\dot H}^2+2H{\ddot H}+6H^2{\dot H}-H^4\right)=0\,.
\end{eqnarray}
To study the evolution of the model, we need to find approximate solutions of Eq.\eqref{Hquad1} in different regimes.  Let us  suppose  that 
\begin{eqnarray} 
6\alpha>> 96\beta H^4\,.
\end{eqnarray}
Then Eq.\eqref{Hquad1} becomes
\begin{eqnarray} 
H^2+6\alpha\left(2H{\ddot H}+6H^2{\dot H}-{\dot H}^2\right)\cong0
\end{eqnarray}
obtaining the well known Starobinsky scalaron mass \cite{Starobinsky1980,vilenkin} 
\begin{eqnarray} 
m^2_R=\frac{1}{6\alpha}
\end{eqnarray}
and the scale factor
\begin{equation}
a(t) \sim \exp \left[{\frac{t}{ \sqrt{6 \alpha}}}\right].
\end{equation}
Considering now the regime   
\begin{eqnarray} 
96\beta H^4 >>   6\alpha\,,
\end{eqnarray}
we have
\begin{eqnarray} 
H^2+96\beta H^4\left(3{\dot H}^2+2H{\ddot H}+6H^2{\dot H}-H^4\right)\cong0\,.
\end{eqnarray}
From the above relation, we get an additional mass term due to the presence of higher order correction \cite{DeLaurentis2015a}
\begin{eqnarray} 
m^2_{\cal G}=\frac{1}{2\sqrt[3]{12 \beta}}\,,
\end{eqnarray}
with the scale factor
\begin{equation}
a(t) \sim \exp\left[{\frac{t}{ \sqrt[6]{96 \beta}}}\right]\,.
\end{equation}
This result shows that we have two inflationary regimes. In general,  the two epochs are determined by the two terms in Eq.\eqref{klein} and  potentials \eqref{pot} that are valid for any $F(R,{\cal G})$ model.

We can conclude that  considering the entire budget of curvature in the effective action, two effective masses, leading  dynamics, are naturally introduced. These ones correspond to two different regimes ruling large scale and very large scale structures.  A part the above toy model, any $F(R,R_{\mu\nu},R_{\mu\nu\alpha\beta})$ can be recast as $F(R,{\cal G})$ thanks to the constraint \eqref{GBinvariant}. This means that $F(R,{\cal G})$ models  are good candidates to figure out  the cosmic evolution at early epochs.

{However, we discussed only a toy model that could become paradigmatic for tracing any cosmic history considering the evolution of curvature contributions. Clearly the more the model is independent from the fine tuning of parameters, the more its naturalness is recovered. In the present model, 
 the coupling constants $\alpha$ and $\beta$ in Eq.\eqref{starotopo}
 should be $\mathcal{O}(1)$ in natural units where $M_{\rm
Planck}=1$. In this regard, one of the   main drawbacks of
Starobinsky model is that it requires very the unnatural hierarchy
$\alpha \gg \beta$ to give rise to a satisfactory inflationary behavior.
In our case,  we have a similar  fine-tuning issue so that both $\alpha$ and $\beta$ have to be adjusted as reported above. The problem can be partially alleviated considering that the double-inflationary regime allows to clearly distinguish between the $R^2$-driven phase with respect to the $R^4\sim{\cal G}^2$-driven phase. In other words, being the energy scales very different, the required fine tuning is less severe.}

\section{The cosmographic approach}
\label{cinque}
The above considerations can be extended towards late cosmic epochs adopting cosmography, that is  a  model independent approach to constrain  cosmological evolution by observational data. It relies on the hypothesis of large-scale homogeneity and isotropy, and combines kinematic parameters via the Taylor expansion of the scale factor. 
The starting point is the definition of the cosmographic parameters \cite{Visser2005}:
\begin{align}
&     \label{hubble parameter}
   H(t)= \frac{1}{a} \frac{da}{dt},\\ 
&  \label{deceleration}
   q(t)= - \frac{1}{a} \frac{d^2 a}{dt^2} \left[\frac{1}{a} \frac{da}{dt}\right]^{-2},\\
&   \label{jerk}
   j(t)=  \frac{1}{a} \frac{d^3 a}{dt^3} \left[\frac{1}{a} \frac{da}{dt}\right]^{-3},\\
& \label{snap}
   s(t)=  \frac{1}{a} \frac{d^4 a}{dt^4} \left[\frac{1}{a} \frac{da}{dt}\right]^{-4},\\
& \label{lerk}
   l(t)= \frac{1}{a} \frac{d^5 a}{dt^5} \left[\frac{1}{a} \frac{da}{dt}\right]^{-5},
\end{align}
where, as above, $H(t)$ is the Hubble parameter, $q(t)$ is the
deceleration parameter which accounts for 
the decelerating or accelerating expansion of the Universe, $j(t)$ and $s(t)$ are the jerk and snap parameters, respectively. The latters may serve as  a geometrical diagnostic of dark energy models \cite{sahni2003statefinder,alam2003exploring}. Finally, $l(t)$ is the lerk parameter also related to high order corrections of the cosmic expansion. It is then useful to relate the Hubble parameter derivative with respect to the cosmic time to the other cosmographic parameters, that is:
\begin{equation}
\dot{H} = -H^2 (1 + q) \ ,
\label{eq: hdot}
\end{equation}
\begin{equation}
\ddot{H} = H^3 (j + 3q + 2) \ ,
\label{eq: h2dot}
\end{equation}
\begin{equation}
\dddot{H} = H^4 \left [ s - 4j - 3q (q + 4) - 6 \right ] \ ,
\label{eq: h3dot}
\end{equation}
\begin{equation}
H^{(4)} = H^5 \left [ l - 5s + 10 (q + 2) j + 30 (q + 2) q + 24 \right ]\ ,
\label{eq: h4dot}
\end{equation}
Using these definitions, it is straightforward to rewrite the Hubble parameter in terms of the cosmographic ones  by expanding it in Taylor series and evaluating cosmographic parameters at the present epoch, that is at redshift $z\simeq 0$ \cite{chiba1998luminosity,capozziello2011comprehensive,Demianski2012}:
\begin{align}
H(z) &= H_0 + \frac{dH}{dz}\Big|_{z=0} z + \frac{1}{2!}\frac{d^2H}{dz^2}\Big|_{z=0} z^2 + \frac{1}{3!}\frac{d^3H}{dz^3}\Big|_{z=0} z^3 +  \cdots                  \nonumber\\
&= H_0\Big[1 + (1+q_0) z +\frac{1}{2}(-q_0^2+j_0)z^2              \nonumber\\
&+ \frac{1}{6}(3q_0^2+3q_0^3-4 q_0 j_0-3 j_0 -s_0)z^3             \nonumber\\
&+ \frac{1}{24}(-12q_0^2-24q_0^3-15q_0^4+32q_0 j_0 +25 q_0^2 j_0  \nonumber\\
&+ 7 q_0 s_0  + 12 j_0-4 j_0^2+ 8 s_0 + l_0) z^4 \Big]  +  \cdots   \label{Hz taylor}
\end{align}

Starting from  the Hubble parameter, it is possible to express  the luminosity distance as a redshift polynomial in term of  cosmographic quantities as  \cite{cattoen2007hubble,capozziello2011comprehensive,Demianski2012,Demianski:2016dsa,Piedipalumbo:2015jya,Piedipalumbo:2013dqa}
\begin{equation}  \label{modulus z}
   d_L(z) = cH_0^{-1}(z + \mathcal{D}_{L,1} z^2 + \mathcal{D}_{L,2} z^3 + \mathcal{D}_{L,3} z^4 + \mathcal{D}_{L,4} z^5) ,
\end{equation}
where the $\mathcal{D}_{L,i}$ quantities are defined as follows
\begin{align}
   \mathcal{D}_{L,1} &= \frac{1}{2}(1-q_0)\,,\\
  \mathcal{D}_{L,2} &= -\frac{1}{6}(1- q_0-3q_0^2+ j_0)\,,\\
   \mathcal{D}_{L,3} &= \frac{1}{24}(2-2 q_0-15q_0^2-15 q_0^3+5 j_0+10 q_0 j_0 +s_0)\,, \\
   \mathcal{D}_{L,4} &= \frac{1}{120}(-6+6q_0+81q_0^2+165q_0^3+105q_0^4    \nonumber\\
   &+10j_0^2-27j_0-110q_0 j_0-105q_0^2 j_0-15q_0 s_0       \nonumber\\
   &-11s_0-l_0) \,.
\end{align}
whereas, the angular diameter distance can be recast as:
\begin{equation}
d_{A}(z) = c H_{0}^{-1} \left ( z +
\mathcal{D}_{A,1}  z^2 + \mathcal{D}_{A,2} \ z^{3} +
\mathcal{D}_{A,3}  z^{4} + \mathcal{D}_{A,4}  z^{5}  \right )\,,
\end{equation}
defining
\begin{align}
\mathcal{D}_{A,1} &= - \frac{1}{2} \left(3 + q_{0}\right)\,, \\
\mathcal{D}_{A,2} &= \frac{1}{6} \left(11 + 7 q_{0} + 3q_{0}^{2} - j_{0} \right)\,, \\
\mathcal{D}_{A,3} &= - \frac{1}{24} (50 + 46 q_{0} + 39
q_{0}^{2} + 15 q_{0}^{3} - 13 j_{0}\nonumber\\ 
& - 10 q_{0} j_{0} - s_{0})\,, \\
\mathcal{D}_{A,4} &= \frac{1}{120} ( 274 + 326 q_{0} + 411
q_{0}^{2} + 315 q_{0}^{3}  \nonumber\\ 
&+ 105 q_{0}^{4}- 210 q_{0} j_{0} - 105
q_{0}^{2} j_{0} - 15 q_{0} s_{0} + \nonumber\\
&- 137 j_{0} + 10 j^{2} - 21 s_{0} - l_{0} )\,.
\end{align}
Cosmological observables  may be used to obtain constraints on the cosmographic parameters by comparing the theoretical predictions with data \cite{aviles2012cosmography}.

\subsection{The {\em y}-redshift cosmography}
\label{duea}

Although cosmography  has been successfully used to constrain the evolution of the late Universe, it is worth  noticing that, at 
at high redshift, i.e. $z>1$, the Taylor expansion does not converge. Therefore cosmography, as introduced above, can be consistently  applied only to low redshift datasets. Nevertheless, in the last decade, a lot of high redshift observations with unprecedented accuracy have been acquired. To be able to use such datasets, one has to solve the convergence problem. The latter may be overcome re-parameterizing the Taylor expansion using the expansion in  $y$-redshift  \cite{cattoen2007hubble}, that is  
\begin{equation}
y = \frac{z}{1+z}\,
\end{equation}
or  $y=1-a(t)$. By definition, the new parameter span the range $[0,1]$, that corresponds to a redshift range $[0, \infty]$, allowing to perform high redshift cosmography.

In the $y$-redshift parameterization, the Hubble function can be re-written as:
\begin{align}\label{eq:hubbble_y}
H(y) &={H_0} \bigg[1+({q_0}+1) y +\frac{1}{2}  \left({j_0}-{q_0}^2+2 {q_0}+2\right)y^2\nonumber\\
&-\frac{1}{6}  \left(4 {j_0} {q_0}-3 {j_0}-3 {q_0}^3+3 {q_0}^2-6 {q_0}+{s_0}-6\right)y^3\nonumber\\
&+\frac{1}{24}  (-4 {j_0}^2+25 {j_0} {q_0}^2-16 {j_0} {q_0}+12 {j_0}+{l_0}-15 {q_0}^4\nonumber\\
&+12 {q_0}^3-12 {q_0}^2+7 {q_0} {s_0}+24 {q_0}-4 {s_0}+24)y^4
 \bigg]\,.
\end{align}

While, luminosity distance can be recast as:
\begin{equation}  \label{eq:dl_y}
   d_L(y) = cH_0^{-1}(y + \mathcal{D}_{L,1}^y  y^2 + \mathcal{D}_{L,2}^y y^3 + \mathcal{D}_{L,3}^y  y^4 + \mathcal{D}_{L,4}^y  y^5) ,
\end{equation}
with
\begin{align}
   \mathcal{D}_{L,1}^y &= \frac{1}{2}(3-q_0)                                         \nonumber\\
  \mathcal{D}_{L,2}^y &= \frac{1}{6}(11-5 q_0+3q_0^2- j_0)                         \nonumber\\
   \mathcal{D}_{L,3}^y &= \frac{1}{24}(50- 26q_0+21 q_0^2-15 q_0^3-7 j_0+10 q_0 j_0 +s_0)  \nonumber\\
  \mathcal{D}_{L,4}^y &= \frac{1}{120}(274-154q_0+141q_0^2-135q_0^3+105q_0^4        \nonumber\\
   &+10j_0^2-47j_0+90q_0 j_0-105q_0^2 j_0-15q_0 s_0                         \nonumber\\
   &+9s_0-l_0) .
\end{align}

Finally, the angular diameter distance is re-written as:
\begin{equation}\label{eq:da_y}
d_{A}(z) = c H_{0}^{-1} \left ( z +
\mathcal{D}_{A,1}^{y}  z^2 + \mathcal{D}_{A,2}^{y} \ z^{3} +
\mathcal{D}_{A,3}^{y}  z^{4} + \mathcal{D}_{A,4}^{y}  z^{5}  \right )\,,
\end{equation}
where we have defined
\begin{align}
\mathcal{D}_{A,1}^y &= - \frac{1}{2} \left(1 + q_{0}\right)\,, \\
\mathcal{D}_{A,2}^y &= \frac{1}{6} (-{j_0}+{q_0} (3 {q_0}-2)-10)\,, \\
\mathcal{D}_{A,3}^y &= \frac{1}{24} \biggl[{j_0} (10 {q_0}-7)+{q_0} (3 {q_0} (7-5 {q_0})+10)\nonumber\\
& +{s_0}-58\biggr]\,, \\
\mathcal{D}_{A,4}^y &= \frac{1}{120} \biggl[10 {j_0}^2+{j_0} (35 (4-3 {q_0}) {q_0}-22)-{l_0}\nonumber\\
& +{q_0} (3 {q_0} (35 ({q_0}-2) {q_0}+22)-15 {s_0}+196) \nonumber\\
& +14 {s_0}-316\biggr]\,.
\end{align}
In the following sections, we will use the $y$-redshift cosmography to obtain constraints on cosmographic parameters, and on generic analytic  $F(R,{\cal G})$ models.

\section{Observational data and methodology}\label{sei}

In order to constrain the cosmographic parameters, we use
measurements  of luminosity distances coming from Supernovae Type Ia  (SNeIa) and Gamma Ray Bursts (GRBs), of $H(z)$, and of Baryonic Acoustic Oscillation (BAO). 
This will allow to set up a large sample of data and then consistently constrain the model.

\subsection{Luminosity distance}\label{sec:SN}
We employ a catalogue of 557  SNeIa in the redshift range $z=[0,1.4]$ (UnionII catalogue \cite{Amanullah+08}), and a list of 109 GRBs given in \cite{Wei10} (the catalogue was compiled using the Amati relation \cite{Amati02,Amati08,Amati09}), of which 50 GRBs at $z<1.4$ and 59 GRBs  distributed in the range of redshift $[0.1, 8.1]$. The observable is the distance modulus $\mu_{obs}$, and its  theoretical counterpart is given  by
\begin{equation}\label{eq:distmod}
\mu_{th} (z) = 5 \log_{10}\hat{d}_{L}(z) + \mu_{0}~,
\end{equation}
where $\hat{d}_{L}(z)\equiv{d}_{L}(z)/(cH_0^{-1})$ and $\mu_{0} = 42.38 - 5 \log_{10}h$, with $h\equiv H_0/100$. It worth  noticing that the above equation holds for both cosmographic parameterizations discussed above. Finally, $\chi^2$ can be computed as
\begin{equation}
-2\log{\cal L}_k({ \bf p})=\chi_k^{2}({ \bf p})=\sum_{i=1}^{N_{obj}}\biggl(\frac{\mu
	_{th}(z_{i}, { \bf p})-\mu _{obs}(z_i)}{\sigma _{\mu}(z_i)}\biggr)^{2},
\end{equation}
where $k=[\mathrm{SN}, \mathrm{GRB}]$, $N_{obj}=[557; 109]$ for SNeIa and GRBs, respectively. Here, $\sigma_{\mu}(z)$ is the error on $\mu _{obs}(z)$. Nevertheless, as well known,  the parameter $\mu_{0}$ encodes $H_0$ and it must be marginalized over. Therefore, the $\chi^{2}$ function can be defined as 
\cite{DiPietro+03, Nesseris+05,	Perivolaropoulos05, Wei10a}:
\begin{equation}
\tilde{\chi}^{2}({ \bf p})= \tilde{A} -
\frac{\tilde{B}^{2}}{\tilde{C}}, \label{eq:chiSN_2}
\end{equation}
where
\begin{align}
\tilde{A} & = \sum_{i=1}^{{N_{obj}}}\biggl(\frac{\mu _{th}(z_{i},{ \bf p}, \mu_{0} =
	0)- \mu _{obs}(z_i)}{\sigma _{\mu}(z_i)}\biggr)^{2}, \\
\tilde{B} & = \sum_{i=1}^{{N_{obj}}}\frac{\mu _{th}(z_{i},{ \bf p}, \mu_{0} =
	0)- \mu _{obs}(z_i)}{\sigma _{\mu}^{2}(z_i)}~, \\
\tilde{C} & = \sum_{i=1}^{{N_{obj}}}\frac{1}{\sigma _{\mu }^{2}(z_i)}~.
\end{align}

\subsection{Expansion rate}
An additional dataset is composed by $30$ uncorrelated measurements of expansion rate, $H(z)$, \cite{Jimenez02,Simon05, Stern10, Moresco12a,Moresco12b, Moresco16a, Moresco15, Zhang14}. Thus, following the prescription in  \cite{Lukovic2016},  the corresponding $\chi ^{2}$ can be defined as
\begin{equation}
-2\log{\cal L}_{\mathrm{H}}({ \bf p})=\chi _{\mathrm{H}}^{2}({ \bf p})=\sum_{i=1}^{30}\biggl(\frac{H(z_{i},{ \bf p})-
	H_{obs,}(z_{i})}{\sigma _{H}(z_{i})}\biggr)^{2}~, \label{eq:chiOHD}
\end{equation}
where $\sigma_{H}(z)$ is the error on $H_{obs}(z)$. 

\subsection{Baryonic Acoustic Oscillations}  

Finally, another additional dataset we are going to use is given by the data from the $6$dFGS \cite{Beutler11}, the SDSS DR7 \cite{Ross15}, the BOSS DR$11$ \cite{Anderson14,Delubac15,Ribera14}, which are also reported in Table I of \cite{Lukovic2016}. As usual, we define the  BAOs observable as : $\hat\Xi\equiv r_d/D_V(z)$; where  $r_{d}$ is the sound horizon at the drag epoch 
and $D_V$ the spherically averaged distance measure \cite{Eisenstein05}
\begin{equation}
D_V(z)\equiv \left[(1+z)^2 d_A^2(z)\dfrac{cz}{H(z)}\right]^{1/3}\,.
\label{eq:D_v}
\end{equation}

For sake of simplicity, since we are adopting a model independent approach, we will set  $r_d=144.57$ Mpc \cite{Planck_dr3}. Finally,  the $\chi^2$ can be straightforwardly computed as
\begin{equation}
-2\log{\cal L}_{\mathrm{BAO}}({ \bf p})=\chi _{\mathrm{BAO}}^{2}({ \bf p})= \sum_{i=1}^{6} \left(\frac{\hat\Xi({ \bf p},z_i)-\Xi_{obs}(z_i)}{\sigma_{\Xi}(z_i)}\right)^2\,,
\end{equation} 
where  $\sigma_{\Xi}(z)$ is the error on ${\Xi}(z)$.

\subsection{The Monte Carlo Markov Chain}

The theoretical counterparts are predicted using Eqs. \eqref{eq:hubbble_y}, \eqref{eq:dl_y}, and \eqref{eq:da_y}, and fit to the aforementioned datasets in order to compute the log-likelihood, $-2\log{\cal L}=\chi^2({ \bf p})$ where  ${\bf p}=[H_0, q_0, j_0, s_0, l_0]$ are 
the parameters of the model. 

We explore the parameter space with a Monte Carlo Markov Chain (MCMC) employing  a Metropolis-Hastings \cite{Metropolis1953, Hastings1970} sampling  algorithm. The step size is adapted to guarantee an acceptance rate between 20\% and 50\% \cite{Gelman1996,Roberts1997}, and the convergence is ensured by the Gelman-Rubin criteria  \cite{Gelman1992}.  Finally,  
the different chains are merged  to constrain the model parameters. Our priors are listed in Table \ref{tab:priors}. 
\begin{table}[!ht]
	\begin{center}
		\begin{tabular}{lcr}
			\hline\hline
			{\bf Parameter} & {\bf Priors} & {\bf Best fit } \\
			\hline\hline\\
			$H_0$  & $[50, 100]$  & $69.84^{+0.83}_{-0.82}$  \\[0.1cm]
			$q_0$  & $[-1, 1]$ & $-0.16^{+0.04}_{-0.05}$ \\[0.1cm]
			$j_0$ & $[-20, 20]$ & $-13.25^{+0.83}_{-0.62}$\\[0.1cm]
			$s_0$ & $[-200, 200]$& $ -138.47^{+13.52}_{-10.21}$\\[0.1cm]
			$l_0$  & $[-200, 200]$ & $36.41^{+18.64}_{-16.80}$\\[0.1cm]
			\hline\hline
		\end{tabular}
		\caption{Parameter space explored by the MCMC algorithm. The first column list the parameters, the second column lists the prior used in our MCMC pipeline, and, finally, the third column lists the best fit values.}\label{tab:priors}
	\end{center}
\end{table}

\begin{figure*}[!ht]
	\centering
	\includegraphics[width=1.9\columnwidth,height=13cm]{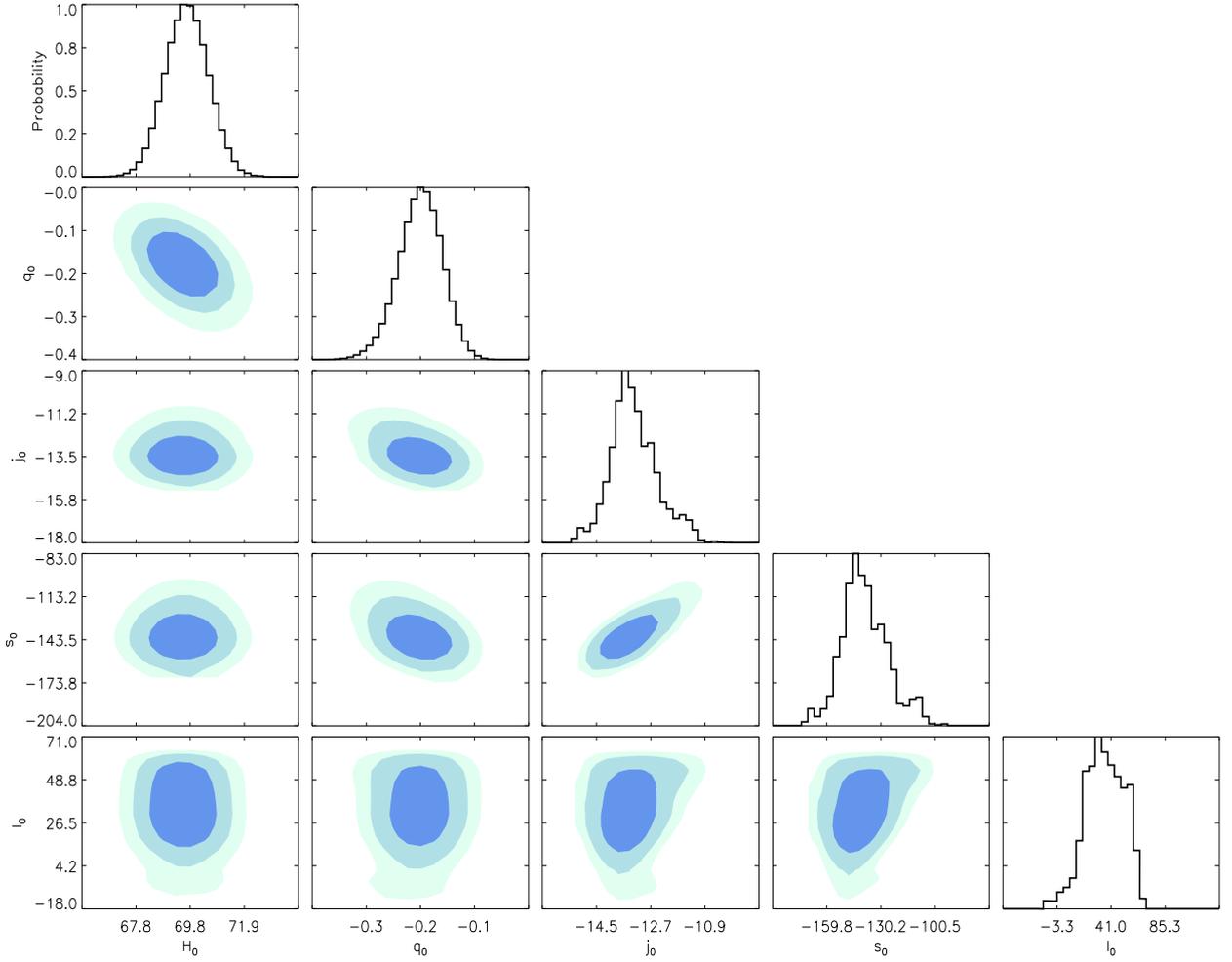}
	\caption{ Two dimensional joint contours of the cosmographic parameters obtained using SNeIa+BAO+H(z)+GRB. 	}\label{fig1}
\end{figure*}

Once our MCMC reaches the convergence, we join  together all likelihoods related to different datasets, $\mathcal{L}(\textbf{p}) = \Pi_i\mathcal{L}_i$ where $i$ indicates the given dataset, to obtain the 2D joint contours represented in Fig. \ref{fig1}, and the best fit parameters summarized in Table \ref{tab:priors}. On one side, our purpose is not just related to find the best fit cosmographic parameters, thus we only show the joint contours at 68\% and 95\% of confidence levels, and then report the corresponding best fit parameters, without analysing  the constraining power of each dataset.  On the other side, having the posterior distribution of the cosmographic parameters, it will allow us to predict, adopting  a Monte Carlo sampling of those distributions, the corresponding values of the theoretical parameters of the $F(R,\mathcal{G})$ gravity models, once we  relate them to the cosmographic ones. Hence, in the next sections we will introduce the  $F(R,\mathcal{G})$ cosmography.

\section{  $F(R,{\cal G})$ Cosmography}\label{sette}

Assuming again a  flat FRW  metric, we will  compute the equations to describe the cosmological evolution in term of cosmographic parameters. Here, to make the text readable, we summarize the whole procedure and report only the final equations needed to translate the constraints from the cosmographic  to the theoretical parameters of $F(R,{\cal G})$ models. For details in  calculations, we refer the reader to Appendix \ref{app:A}. 

Starting from Eqs. \eqref{FRWa} and \eqref{FRWa2}, we assume that  $F(R,\mathcal{G})$ is expandable in Taylor series, that is:

\begin{align}
&F(R,\mathcal{G})\approx  F(R_0,\mathcal{G}_0)+F_{\mathcal{G}}(R_0,\mathcal{G}_0) \mathcal{G}+\frac{1}{2} F_{\mathcal{G}\mathcal{G}}(R_0,\mathcal{G}_0) \mathcal{G}^2+ \nonumber\\
& R \left(F_{R}(R_0,\mathcal{G}_0) + F_{R\mathcal{G}}(R_0,\mathcal{G}_0) \mathcal{G}+\frac{1}{2} F_{R\mathcal{G}\mathcal{G}} (R_0,\mathcal{G}_0)\mathcal{G}^2\right)+\nonumber\\
& R^2 \biggl(\frac{F_{RR}(R_0,\mathcal{G}_0)}{2}+\frac{1}{2} F_{RR\mathcal{G}}(R_0,\mathcal{G}_0) \mathcal{G}+\nonumber\\
&\frac{1}{4} F_{RR\mathcal{G}\mathcal{G}}(R_0,\mathcal{G}_0) \mathcal{G}^2\biggr)\,.   
\end{align}
Supposing that the models may be well approximated by the second order Taylor expansion in $(R - R_0)$ and $(\mathcal{G}-\mathcal{G}_0)$, one can  evaluate the main equations describing the cosmological evolution, that is  \eqref{eq:Hsqr} and \eqref{eq:Hdot} at the present day, using the definition of the Ricci and Gauss-Bonnet scalars in term of the cosmographic parameters. To this end, we need to consider the time derivative of \eqref{eq:Hdot} and time derivatives of $R$ and  $\cal G$. The result is the system of equations in \eqref{eq:Hddot}-\eqref{eq:tayass}. Then, we have to  solve the  system of equations with respect to the present day values of $F(R,\mathcal{G})$ and its derivatives up to the second order of the expansion. A reasonable approximation is to neglect all term beyond $R^3$ in the Taylor expansion, which means to set the following conditions:
\begin{align}
\label{eq:assumptions1}
F_{\mathcal{G}\mathcal{G}}(R_0,\mathcal{G}_0) = F_{R\mathcal{G}}(R_0,\mathcal{G}_0) = F_{R\mathcal{G}\mathcal{G}}(R_0,\mathcal{G}_0) = 0 ,
\end{align}
and
\begin{equation}
\label{eq:assumptions2}
    F_{RR\mathcal{G}}(R_0,\mathcal{G}_0) =F_{RR\mathcal{G}\mathcal{G}}(R_0,\mathcal{G}_0) =0.
\end{equation}

Finally, since we want to  recover GR at  lower order, we set the prior:
\begin{equation}
\label{eq:assumptions3}
F_{R}(R_0,\mathcal{G}_0) = 1 \,.
\end{equation}
In such a way, one  obtains\,:
\begin{equation}
\label{eq: fRG_result1}
\frac{F(R_0,\mathcal{G}_0)}{H_0^2 } =  \frac{{\cal{P}}_1(q_0,
j_0, s_0, l_0) + {\cal{P}}_2(q_0, j_0, s_0, l_0, \Omega_M)}{{\cal{R}}(q_0, j_0, s_0,
l_0)} \ ,
\end{equation}

\begin{equation}
\frac{F_{RR}(R_0,\mathcal{G}_0)}{\left ( 6 H_0^2 \right )^{-1}} =  \frac{{\cal{P}}_3(q_0,
j_0, s_0, l_0) + {\cal{P}}_4(q_0, j_0, s_0, l_0, \Omega_M)}{{\cal{R}}(q_0, j_0, s_0,
l_0)} \ ,
\label{eq: fRG_result2}
\end{equation}

\begin{equation}
\frac{F_{R\mathcal{G}\mathcal{G}}(R_0,\mathcal{G}_0)}{\left ( 48 H_0^{4} \right )^{-1}} =  \frac{{\cal{P}}_5(q_0,
j_0, s_0, l_0) + {\cal{P}}_6(q_0, j_0, s_0, l_0, \Omega_M)}{{\cal{R}}(q_0, j_0, s_0,
l_0)} \ ,
\label{eq: fRG_result3}
\end{equation}
where we have defined the  auxiliary functions: ${\cal{P}}_1(q_0, j_0, s_0, l_0)$, ${\cal{P}}_2(q_0, j_0, s_0, l_0, \Omega_m)$, ${\mathcal{P}}_3(q_0, j_0, s_0, l_0)$, ${\mathcal{P}}_4 (q_0, j_0, s_0, l_0, \Omega_m)$, ${\mathcal{P}}_5 (q_0, j_0, s_0, l_0) $, ${\mathcal{P}}_6 (q_0, j_0, s_0, l_0, \Omega_m)$, and ${\mathcal{R}}(q_0, j_0, s_0, l_0)$, which are reported in  Appendix \ref{app:A}, see Eqs. \eqref{eq:pol1}-\eqref{eq:Rpol}. 

Assumptions in Eqs. \eqref{eq:assumptions1}-\eqref{eq:assumptions3} are made in order to: {\em (i)} analyze the $F(R,\mathcal{G})$ model without any term contributing as a cosmological constant, thus we force the  zero-order of the Taylor expansion to be zero; {\em (ii)} have only the Ricci scalar as first term of the Taylor expansion  to recover the GR. In such a way we are focusing on the contributions given by higher-order terms in $R$ and $\mathcal{G}$ to the cosmological evolution and, also, we are avoiding  to consider  extra equations in our systems which would lead to the introduction of extra cosmographic parameters which would decrease the constraining power of the dataset.

\begin{figure*}[!ht]
	\centering
	\includegraphics[width=2.0\columnwidth,height=13cm]{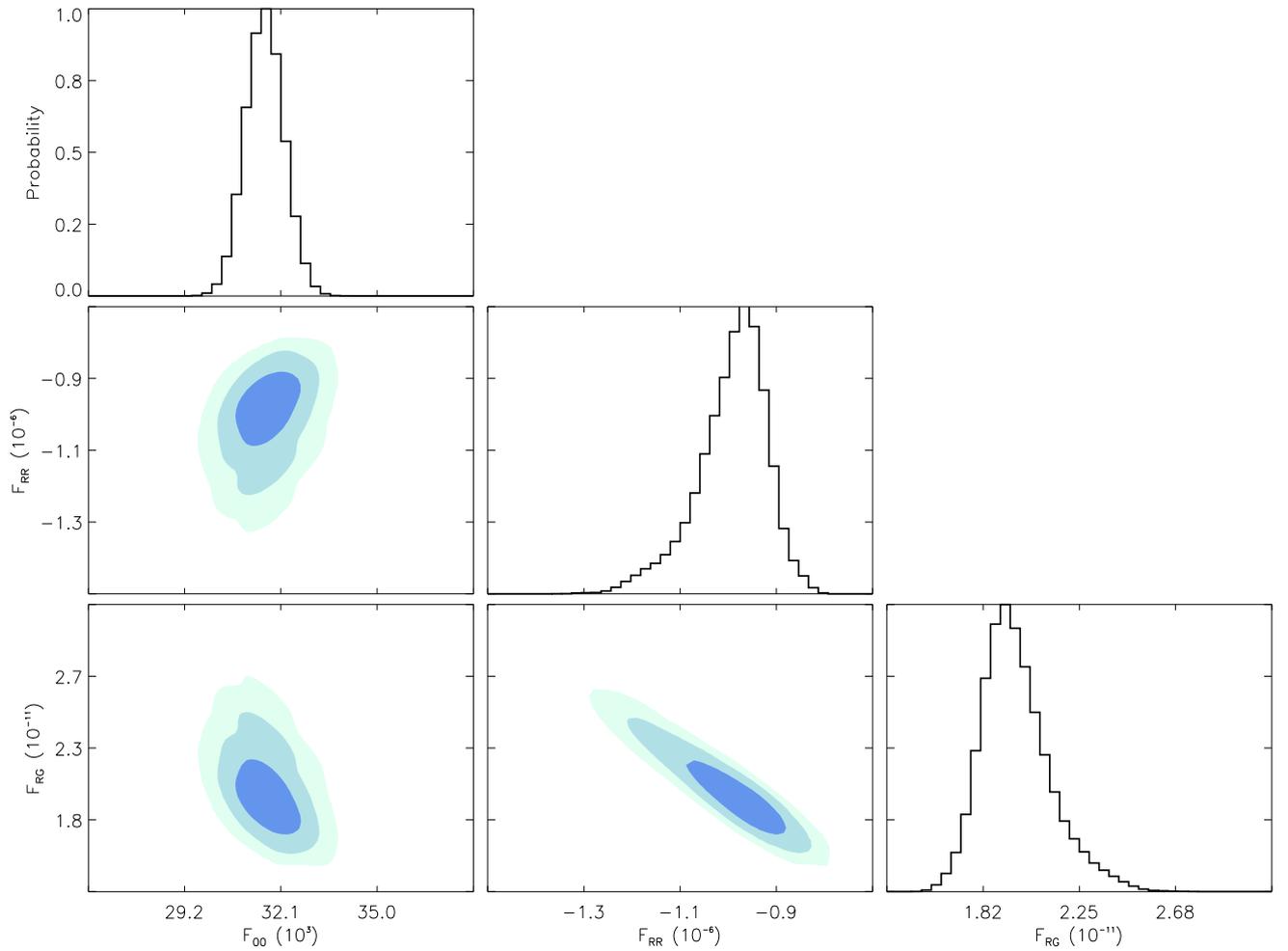}
	\caption{Two dimensional joint contours of $F(R,\mathcal{G})$ model's parameters corresponding to a Monte Carlo sampling of the posterior distribution in Fig.\ref{fig1}.}\label{fig2}
\end{figure*}

Eqs. \eqref{eq: fRG_result1}-\eqref{eq: fRG_result3} allow to use  constraints on the cosmographic parameters, given in Table \ref{tab:priors}, to bounds the derivatives of the Taylor expansion
 of  $F(R,\mathcal{G})$ model. Thus, we carried out 1000 Monte
 Carlo simulations randomly choosing the values of the cosmographic 
 parameters from their posterior distribution, and deriving the 2D contours and their best values. Final results are summarized in Table \ref{tab:bestfitFRG}, and in Figure \ref{fig2}. 
 \begin{table}[!ht]
	\begin{center}
		\begin{tabular}{lr}
			\hline\hline
			{\bf Parameter} &{\bf Best fit } \\
			\hline\hline\\[-0.1cm]
			$F_{00}\,(10^{3})$   & $31.73^{+0.58}_{-0.57}$ \\[0.1cm]
			$F_{RR}\,(10^{-6})$   & $-0.95^{+0.05}_{-0.08}$ \\[0.1cm]
			$F_{RG}\,(10^{-11})$   & $1.97^{+0.17}_{-0.12}$\\[0.1cm] 
			\hline\hline
		\end{tabular}
		\caption{Best fit parameters of the $F(R,\mathcal{G})$ model corresponding to the one dimensional posterior distribution in Fig. \ref{fig2}.}\label{tab:bestfitFRG}
	\end{center}
\end{table}
 
 Results deserve some  comments. As it is possible to see from the bounded values, higher order Gauss-Bonnet terms  do not affect the late-time cosmological evolution. It is fully driven by the $R^2$ term of the Taylor expansion. In other words, the $F(R,\mathcal{G})$ theory reduces to  $F(R)$ models when late time evolution is considered. 
 
This output is rather expected because the Gauss-Bonnet invariant, scaling as quadratic gravity, works very well in  high curvature regimes  as those of  primordial epochs. As shown above, a term like ${\cal G}^2$ scales as $R^4$ and then it is effective at very high energies naturally producing inflationary behaviors. Clearly, it must be negligible at  late time. 

Furthermore,  bounds on the second derivative with respect to the Ricci scalar is consistent with 
constraints obtained  in  $F(R)$ gravity as shown, for example, in \cite{2008PhRvD..78f3504C,Orlandof(R)}. This may be considered as a self-consistent test for the  procedure. In a different perspective, it seems that cosmological evolution is efficiently described by two fields  at early epoch while one field is sufficient to describe late epochs. In other words,  the Klein-Gordon dynamics,  given by \eqref{klein}, turns on or turns off the scalar fields (related to $R$ and $\cal G$) according to the scale.

\section{Discussion and Conclusions}
\label{Conclusions}

One of the main goals of modern cosmology is to construct  self-consistent models capable of tracking cosmic history from early to late epochs. According to this program, inflation, dark energy and dark matter issues have to be consider in order to describe evolution at any era. 

Being, up to now, the dark side so elusive because no final matter candidates have been detected  at fundamental level, improving geometric sector seems a reliable approach. Adding further curvature and torsion invariants is a paradigm supported by effective gravitational theories formulated in curved space-time: these terms emerge in the action as soon as one faces the problem to regularize and renormalize the theory.  Specifically, terms containing $R^2$, $R_{\mu\nu}R^{\mu\nu}$, $R_{\mu\nu\alpha\beta}R^{\mu\nu\alpha\beta}$, eventually constrained by the Gauss-Bonnet invariant $\cal G$, have to be considered in any approach aimed to obtain a theory of gravity renormalized at one-loop level \cite{Birrell}.

In this perspective, any cosmological model which wants to take advantage from this extended theories with respect to GR has to consider such invariants. 

Models like $F(R,{\cal G})$, in principle, take into account  second-order curvature invariants that give rise to fourth-order field equations by varying with respect to the metric.  Considering $\cal G$ as a constraint  for the other terms means that such an action is capable of representing all the effective degrees of freedom related to second order curvature invariants and, furthermore, one is dealing with an effective theory with two scalar fields.

In this paper, we discussed $F(R,{\cal G})$ models at early and late time epochs, that means at ultraviolet and infrared regimes.  

The main result at  early epochs is that  a double inflation, depending on $\cal G$ and $R$, can be naturally achieved. This feature is  very important  to produce large and very large structures and to give a gateway mechanism to regularize the inflation (see for example \cite{ACLO}).

A cosmographic approach has been adopted at late epochs for redshift $z\rightarrow 0$. Without choosing specific $F(R,{\cal G})$ models, we investigated if GR extensions due to $R$ and $\cal G$ affect dynamics at recent time. Among the priors, we imposed that GR has to be  recovered and cosmographic parameters, in the observed ranges, are restored. In the approximation, we excluded adding further cosmographic parameters due to the order of approximation. The result is that $F(R)$ corrections have to be included while $\cal G$ and its functions  are negligible at late time.  The interpretation of this fact is quite straightforward: dark energy regime can be restored just considering an effective field (i.e. the  field related to $F(R)$) and two scalar fields are not necessary at late epochs. 

{This result is coherent with the following fact. It is well known that terms like ${\cal G}^2$ can give rise to phantom solutions in the limit $t\to \infty$. In this case,   the de Sitter phase  is asymptotically unstable as reported in literature (see
\cite{Nojiri2007,Nojiri2006,Bazeia2007,Cognola2007,Barrow2007,Bazeia2008,Goheer2009a,Goheer2009b,Mohsenir2010,Sadjadi2010,Alimohammadi2009,Boehmer2009,Uddin2009,Zhou2009,Ivanov2012,Nojiri2010,KumarSanyal2011,Grandal2012,Grandal2013,Benetti1,Benetti2} and references therein).  In the present study, however, this problem is overcome because terms like ${\cal G}^2$ dominates at $t\to 0$, leading the first  inflationary phase while they  decay at late epochs: here $R$ terms are dominating and then the de Sitter solution results stable and phantom-free.  From an observational viewpoint, this statement is supported by the above contour plots based on the reported data sets: cosmographic results exclude the contributions of $\cal G$ terms into late time dynamics and, as a consequence, phantom instability is also excluded from the ranges of parameters. In other words, the best fit values in Table \ref{tab:priors}
for the cosmographic parameters and in Table \ref{tab:bestfitFRG} for $F(R,{\cal G})$ parameters exclude phantom behaviors.}

Clearly, what we discussed here is a coarse-grained approach that deserves further investigations and more refined studies. This will be the topic of a next paper.

\section*{Acknowledgements}
This article is supported by COST Action CA15117 "Cosmology and Astrophysics Network for Theoretical Advances and Training Action" (CANTATA) of the COST (European Cooperation in Science and Technology). IDM  is supported by the grant ``The Milky Way and Dwarf Weights with Space Scales" funded by University of Torino and Compagnia di S. Paolo (UniTO-CSP).  IDM also acknowledge partial support from the INFN grant InDark. M.D.L. acknowledges INFN Sez. di Napoli (Iniziative Specifica TEONGRAV). S.C. and M.D.L. acknowledge INFN Sez. di Napoli (Iniziativa Specifica QGSKY).

\begin{widetext}
\appendix
\section{Detailed calculations for  {\em F(R,{\cal G})} cosmography}\label{app:A}

In the FRW background in Eq. \eqref{metric} with a perfect fluid equation of state for ordinary matter, the field equations for $F(R,{\cal G})$ gravity are given by
\begin{eqnarray}\label{FRWa}
\left(\frac{{\ddot a}a-{\dot a}^2}{a^2}\right)F_R&=&-\frac{\kappa}{2}(p^{(m)}+\rho^{(m)}+\frac{1}{
2}\biggl[\left(\frac{\dot a}{a} \right){\dot F}_R-\ddot{F}_R+4 \left(\frac{\dot a}{a}\right)^3\dot{F}_{\cal G}-8 \left(\frac{\dot a}{a}\right) \left(\frac{{\ddot a}a-{\dot a}^2}{a^2}\right)\dot{F}_{\cal G}-4\left(\frac{\dot a}{a}\right)^2\ddot{F}_{\cal G}\biggr],\\
\nonumber\\
\label{FRWa2}
F_R \left(\frac{\dot a}{a}\right)^2&=&{\frac{\kappa}{3}}\rho^{(m)}+\frac{1}{6}\biggl[R\,F_R-F(R,{\cal G})-6\left(\frac{\dot a}{a}\right)\dot{F}_R+{\cal G}F_{\cal G}-24\left(\frac{\dot a}{a}\right)^3\dot{F}_{\cal G}\biggr]\,,
\end{eqnarray}
where $\rho^{(m)}$ and $p^{(m)}$ are the energy density and pressure of ordinary matter, respectively, and the overdot denotes a derivative with respect to the time coordinate, $t$.

Assuming that $\rho^{(m)} = 3 H_0^2 \Omega_m a^{-3}(t)$ and $p^{(m)} = 0$, and expanding the time derivatives, the above equations \eqref{FRWa} and \eqref{FRWa2} become:
\begin{align}
F_{R} H^2 & =\frac{1}{6} \left(-6 H \left(4 H^2 \left(\dot{\mathcal{G}} F_{\mathcal{G}\mathcal{G}}+\dot{R} F_{R\mathcal{G}}\right)+\dot{\mathcal{G}} F_{R\mathcal{G}}+\dot{R} F_{RR}\right)+\mathcal{G} F_{\mathcal{G}}+R F_{R}-F\right)+\frac{{H_0}^2 \kappa {\Omega_m}}{a^3}\,,\label{eq:Hsqr}
\end{align}
\begin{align}
    2  F_{R}\dot{H} =& -\ddot{\mathcal{G}} F_{R\mathcal{G}}+H \left(\dot{\mathcal{G}} \left(F_{R\mathcal{G}}-8 \dot{H} F_{\mathcal{G}\mathcal{G}}\right)+\dot{R} \left(F_{RR}-8 \dot{H} F_{R\mathcal{G}}\right)\right)+4 H^3 \left(\dot{\mathcal{G}} F_{\mathcal{G}\mathcal{G}}+\dot{R} F_{R\mathcal{G}}\right)\nonumber\\
& -\dot{\mathcal{G}} \left(\dot{\mathcal{G}} F_{R\mathcal{G}\mathcal{G}}+2 \dot{R} F_{RR\mathcal{G}}\right)-4 H^2 \left(\ddot{\mathcal{G}} F_{\mathcal{G}\mathcal{G}}+2 \dot{\mathcal{G}} \dot{R} F_{R\mathcal{G}\mathcal{G}}+\dot{\mathcal{G}}^2 F_{\mathcal{G}\mathcal{G}\mathcal{G}}+\ddot{R} F_{R\mathcal{G}}+\dot{R}^2 F_{RR\mathcal{G}}\right)\nonumber\\
&   -\ddot{R} F_{RR}-\dot{R}^2 F_{RRR}-\frac{3 H_0^2 \kappa \Omega_m}{a^3} \label{eq:Hdot}
\end{align}{}

For the sake of simplicity, we have used the following abbreviations: $F\equiv F(R,{\cal G})$, $F_x\equiv F_x(R,{\cal G})$ where $x$ can be the Ricci or the Gauss-Bonnet scalar, as well as any combination of them indicating derivatives with respect to these variables and, finally, we have also defined $a\equiv a(t)$, $H\equiv H(t)$, $R\equiv R(t)$, and ${\cal G}(t)\equiv {\cal G}$. 
To complete the dynamical system, we have to consider constraints coming from the definitions of $R$  and $\cal G$, that is 
\begin{equation}
\label{laglag}
 R = 6 \left[\frac{\ddot a}{a}+\left(\frac{\dot a}{a}\right)^2
 \right]\,,\quad \text{
 and} \quad
   {\cal G} = 24 \left(\frac{\ddot{a}\dot{a}^2}{a^3}\right)\,.
   \end{equation}
These last equations are Lagrange multipliers that constrain dynamics. See \cite{FelixOdin} for details.
In order to get  a close system of equations useful for cosmographic analysis, let us differentiate  Eq.\eqref{eq:Hdot} with respect to $t$. We obtain: 
\begin{align}\label{eq:Hddot}
2  F_{R}^2 \ddot{H} = & 4 H^3  \biggl[\left(F_{\mathcal{G}\mathcal{G}\mathcal{G}} F_{R}-F_{\mathcal{G}\mathcal{G}} F_{R\mathcal{G}}\right) \dot{\mathcal{G}}^2-\dot{R} \left(F_{R\mathcal{G}}^2-2 F_{R} F_{R\mathcal{G}\mathcal{G}}+F_{\mathcal{G}\mathcal{G}} F_{RR}\right) \dot{\mathcal{G}}+\ddot{\mathcal{G}} F_{\mathcal{G}\mathcal{G}} F_{R}+\nonumber\\
& F_{R\mathcal{G}} \left(\ddot{R} F_{R}-\dot{R}^2 F_{RR}\right)+\dot{R}^2 F_{R} F_{RR\mathcal{G}}\biggr] +4 H^2  \biggl[\left(F_{\mathcal{G}\mathcal{G}\mathcal{G}} F_{R\mathcal{G}}-F_{\mathcal{G}\mathcal{G}\mathcal{G}\mathcal{G}} F_{R}\right) \dot{\mathcal{G}}^3+\nonumber\\
& \dot{R} \left(2 F_{R\mathcal{G}} F_{R\mathcal{G}\mathcal{G}}-3 F_{R} F_{R\mathcal{G}\mathcal{G}\mathcal{G}}+F_{\mathcal{G}\mathcal{G}\mathcal{G}} F_{RR}\right) \dot{\mathcal{G}}^2+\biggl(2 F_{R\mathcal{G}\mathcal{G}} F_{RR} \dot{R}^2+F_{R\mathcal{G}} F_{RR\mathcal{G}} \dot{R}^2-3 F_{R} F_{RR\mathcal{G}\mathcal{G}} \dot{R}^2+\nonumber\\
& \ddot{R} F_{R\mathcal{G}}^2+3 \dot{H} F_{\mathcal{G}\mathcal{G}} F_{R}+\ddot{\mathcal{G}} \left(F_{\mathcal{G}\mathcal{G}} F_{R\mathcal{G}}-3 F_{\mathcal{G}\mathcal{G}\mathcal{G}} F_{R}\right)-3 \ddot{R} F_{R} F_{R\mathcal{G}\mathcal{G}}\biggr) \dot{\mathcal{G}}-G^{(3)} F_{\mathcal{G}\mathcal{G}} F_{R}+\nonumber\\
& 3 \dot{H} \dot{R} F_{R} F_{R\mathcal{G}}-R^{(3)} F_{R} F_{R\mathcal{G}}-3 \dot{R} \ddot{\mathcal{G}} F_{R} F_{R\mathcal{G}\mathcal{G}}+\dot{R} \ddot{\mathcal{G}} F_{\mathcal{G}\mathcal{G}} F_{RR}+\dot{R} \ddot{R} F_{R\mathcal{G}} F_{RR}-3 \dot{R} \ddot{R} F_{R} F_{RR\mathcal{G}}+\nonumber\\
& \dot{R}^3 F_{RR} F_{RR\mathcal{G}}-\dot{R}^3 F_{R} F_{RRR\mathcal{G}}\biggr] +  \biggl[\biggl(F_{R\mathcal{G}} F_{R\mathcal{G}\mathcal{G}}-F_{R} F_{R\mathcal{G}\mathcal{G}\mathcal{G}}\biggr) \dot{\mathcal{G}}^3+\dot{R} \dot{\mathcal{G}}^2 \biggl(F_{R\mathcal{G}\mathcal{G}} F_{RR}+2 F_{R\mathcal{G}} F_{RR\mathcal{G}}-\nonumber\\
& 3 F_{R} F_{RR\mathcal{G}\mathcal{G}}\biggr) +\biggl(-8 F_{\mathcal{G}\mathcal{G}} F_{R} \dot{H}^2+F_{R} F_{R\mathcal{G}} \dot{H}+\ddot{\mathcal{G}} \left(F_{R\mathcal{G}}^2-3 F_{R} F_{R\mathcal{G}\mathcal{G}}\right)+\ddot{R} F_{R\mathcal{G}} F_{RR}-3 \ddot{R} F_{R} F_{RR\mathcal{G}}+\nonumber\\
& 2 \dot{R}^2 F_{RR} F_{RR\mathcal{G}}+\dot{R}^2 F_{R\mathcal{G}} F_{RRR}-3 \dot{R}^2 F_{R} F_{RRR\mathcal{G}}\biggr) \dot{\mathcal{G}}+\dot{R} \ddot{R} F_{RR}^2-8 \dot{H}^2 \dot{R} F_{R} F_{R\mathcal{G}}-G^{(3)} F_{R} F_{R\mathcal{G}}+\nonumber\\
& \dot{H} \dot{R} F_{R} F_{RR}-R^{(3)} F_{R} F_{RR}+\dot{R} \ddot{\mathcal{G}} F_{R\mathcal{G}} F_{RR}-3 \dot{R} \ddot{\mathcal{G}} F_{R} F_{RR\mathcal{G}}-3 \dot{R} \ddot{R} F_{R} F_{RRR}+\dot{R}^3 F_{RR} F_{RRR}-\nonumber\\
& \dot{R}^3 F_{R} F_{RRRR}\biggr] +H \biggl\{\biggl[\dot{\mathcal{G}} \left(8 \dot{H} F_{\mathcal{G}\mathcal{G}}-F_{R\mathcal{G}}\right)+\dot{R} \left(8 \dot{H} F_{R\mathcal{G}}-F_{RR}\right)\biggr] \left(\dot{\mathcal{G}} F_{R\mathcal{G}}+\dot{R} F_{RR}\right)+\nonumber\\
& F_{R} \biggl[\biggl(F_{R\mathcal{G}\mathcal{G}} \dot{\mathcal{G}}^2+\left(2 \dot{R} F_{RR\mathcal{G}}-8 \ddot{H} F_{\mathcal{G}\mathcal{G}}\right) \dot{\mathcal{G}}+ \left(\ddot{\mathcal{G}}-8 \dot{R} \ddot{H}\right) F_{R\mathcal{G}}+\ddot{R} F_{RR}-\nonumber\\
&16 \dot{H} \left(F_{\mathcal{G}\mathcal{G}\mathcal{G}} \dot{\mathcal{G}}^2+2 \dot{R} F_{R\mathcal{G}\mathcal{G}} \dot{\mathcal{G}}+\ddot{\mathcal{G}} F_{\mathcal{G}\mathcal{G}}+\ddot{R} F_{R\mathcal{G}}+\dot{R}^2 F_{RR\mathcal{G}}\right)+\dot{R}^2 F_{RRR}\biggr) +\frac{9 H_0^2 \kappa \Omega_m}{a^3}\biggr]\biggr\}+\nonumber\\
&  \frac{3 H_0^2 \kappa \Omega_m}{a^3 } \left(\dot{\mathcal{G}} F_{R\mathcal{G}}+\dot{R} F_{RR}\right)    
\end{align}

In addition to the previous equations, we may define the Ricci and Gauss-Bonnet scalars and their derivatives with respect to the cosmic time $t$ as function of the Hubble parameter $H$ and its derivatives. Thus, we have 
\begin{align}
 & R = 6 \left(2H^{2}+\dot H \right)\,, \label{eq:R}  \\
 & \dot{R} = 6 \left(\ddot{H}+4 H \dot{H}\right) \label{eq:Rdot}\,,\\
 & \ddot{R} = 6 \left(H^{(3)}+4 H \ddot{H}+4 \dot{H}^2\right)\label{eq:Rddot}\,, \\
 & R^{(3)} = 6 \left(H^{(4)}+4 H H^{(3)}+12 \dot{H} \ddot{H}\right)\label{eq:R3dot}\,,  \\
\end{align}
and
\begin{align} 
&{\cal G} = 24H^{2} \left( H^{2}+\dot H \right)\,, \label{eq:G}\\
 & \dot{{\cal G}} = 48 H \dot{H} \left(\dot{H}+H^2\right)+24 H^2 \left(\ddot{H}+2 H \dot{H}\right) \label{eq:Gdot}\,,\\
  & \ddot{{\cal G}} = 48 \dot{H}^2 \left(\dot{H}+H^2\right)+48 H \left(\dot{H}+H^2\right) \ddot{H}+\nonumber\\
  & \qquad 96 H \dot{H} \left(\ddot{H}+2 H \dot{H}\right)+24 H^2 \left(H^{(3)}+2 H \ddot{H}+2 \dot{H}^2\right) \label{eq:Gddot}\,,\\
 & {\cal G}^{(3)} = 48 H^{(3)} H \left(\dot{H}+H^2\right)+144 H \ddot{H} \left(\ddot{H}+2 H \dot{H}\right)+\nonumber\\
  & \qquad 144 \dot{H} \left(\dot{H}+H^2\right) \ddot{H}+144 \dot{H}^2 \left(\ddot{H}+2 H \dot{H}\right)+\nonumber\\
  & \qquad 144 H \dot{H} \left(H^{(3)}+2 H \ddot{H}+2 \dot{H}^2\right)+\nonumber\\
  & \qquad 24 H^2 \left(H^{(4)}+2 H H^{(3)}+6 \dot{H} \ddot{H}\right) \label{eq:G3dot}\,,
\end{align}

Let us now suppose that the $F(R,\mathcal{G})$-Lagrangian may be well approximated by its second order Taylor expansion in $(R-R_0)$ and $(\mathcal{G}-\mathcal{G}_0)$.  We set:
\begin{align}
F(R,\mathcal{G})\approx&  F(R_0,\mathcal{G}_0)+F_{\mathcal{G}}(R_0,\mathcal{G}_0) \mathcal{G}+\frac{1}{2} F_{\mathcal{G}\mathcal{G}}(R_0,\mathcal{G}_0) \mathcal{G}^2+ R \left(F_{R}(R_0,\mathcal{G}_0) + F_{R\mathcal{G}}(R_0,\mathcal{G}_0) \mathcal{G}+\frac{1}{2} F_{R\mathcal{G}\mathcal{G}} (R_0,\mathcal{G}_0)\mathcal{G}^2\right)+\nonumber\\
& R^2 \biggl(\frac{F_{RR}(R_0,\mathcal{G}_0)}{2}+\frac{1}{2} F_{RR\mathcal{G}}(R_0,\mathcal{G}_0) \mathcal{G}+\nonumber\frac{1}{4} F_{RR\mathcal{G}\mathcal{G}}(R_0,\mathcal{G}_0) \mathcal{G}^2\biggr)\,.  
\end{align}
We can make further assumptions to reduce the complexity of the problem. We want our model to be an extension  of GR then we retain  corrections related to the Gauss-Bonnet invariant up to  $R^3$. The reason for this choice is to simplify the problem in view of  obtaining analytic solutions, and to avoid the introduction of new cosmographic parameters beyond the parameter $l$, which would be needed if higher order correction to GR are taken into account. All those assumptions can be  translated is the following bounds:
\begin{align}
 F_{\mathcal{G}\mathcal{G}}(R_0,\mathcal{G}_0) = F_{R\mathcal{G}}(R_0,\mathcal{G}_0) = F_{R\mathcal{G}\mathcal{G}}(R_0,\mathcal{G}_0)= F_{RR\mathcal{G}}(R_0,\mathcal{G}_0) =F_{RR\mathcal{G}\mathcal{G}}(R_0,\mathcal{G}_0) =0 ,
\end{align}
\begin{equation}
F_{R}(R_0,\mathcal{G}_0) = 1 \label{eq:tayass}\ ,
\end{equation}

Inserting Eq. \eqref{eq:R}-\eqref{eq:tayass} in the Eqs.
\eqref{eq:Hsqr}-\eqref{eq:Hddot}, one gets:
\begin{align}\label{eq:T0}
& (12\mathcal{D})^{-1}\biggl[\frac{{H_0}^2 k {\Omega_m}}{ a^3}- \biggl(2 {F_{00}}+12 H \left({F_{R\mathcal{G}}} \dot{\mathcal{G}}+{F_{RR}} \dot{R}\right)+12 H^2 ({F_{R\mathcal{G}}} G+{F_{RR}} R+1)-\nonumber\\
&R (2 {F_{R\mathcal{G}}} G+{F_{RR}} R)+48 {F_{R\mathcal{G}}} H^3 \dot{R}\biggr)\biggr]=0\,,
\end{align}
\begin{align}
&\dot{H}-(2 \mathcal{D})^{-1}\biggl(H \left(\dot{R} \left({F_{RR}}-8 {F_{R\mathcal{G}}} H'\right)+{F_{R\mathcal{G}}} \dot{\mathcal{G}}\right)-{F_{R\mathcal{G}}} \ddot{\mathcal{G}}-4 {F_{R\mathcal{G}}} H^2 \ddot{R}+\nonumber\\
&4 {F_{R\mathcal{G}}} H^3 \dot{R}-{F_{RR}} \ddot{R}-\frac{3 {H_0}^2 k {\Omega_m}}{a^3 }\biggr) = 0 \,,
\end{align}
\begin{align}
&\ddot{H} - (2  \mathcal{D}^2)^{-1}\biggl\{ \biggl[{F_{R\mathcal{G}}}^2 \dot{\mathcal{G}} \ddot{\mathcal{G}}+{F_{R\mathcal{G}}} \biggl(\mathcal{D} \left(H' \left(\dot{\mathcal{G}}-8 H' \dot{R}\right)-G^{(3)}\right)+{F_{RR}} \ddot{\mathcal{G}} \dot{R}+\nonumber\\
&{F_{RR}} \dot{\mathcal{G}} \ddot{R}\biggr)+{F_{RR}} \left(H' \dot{R}-R^{(3)}\right) \mathcal{D}+{F_{RR}}^2 \dot{R} \ddot{R}\biggr]+4 {F_{R\mathcal{G}}}  H^2 \biggl(\ddot{R} \bigl({F_{R\mathcal{G}}} \dot{\mathcal{G}}+\nonumber\\
& {F_{RR}} \dot{R}\bigr)+3 H' \dot{R} \mathcal{D}-R^{(3)} \mathcal{D}\biggr)+4 {F_{R\mathcal{G}}}  H^3 \left(\ddot{R} \mathcal{D}-\dot{R} \left({F_{R\mathcal{G}}} \dot{\mathcal{G}}+{F_{RR}} \dot{R}\right)\right)+\nonumber\\
& H \biggl(\mathcal{D} \left( \left(\ddot{R} \left({F_{RR}}-16 {F_{R\mathcal{G}}} H'\right)+{F_{R\mathcal{G}}} \ddot{\mathcal{G}}-8 {F_{R\mathcal{G}}} H'' \dot{R}\right)+9 {H_0}^2 k {\Omega_m}\right)+ \nonumber\\
& \left({F_{R\mathcal{G}}} \dot{\mathcal{G}}+{F_{RR}} \dot{R}\right) \left(-\dot{R} \left({F_{RR}}-8 {F_{R\mathcal{G}}} H'\right)-{F_{R\mathcal{G}}} \dot{\mathcal{G}}\right)\biggr)+\frac{3 {H_0}^2 k {\Omega_m}}{a^3} \left({F_{R\mathcal{G}}} \dot{\mathcal{G}}+{F_{RR}} \dot{R}\right)\biggr\} = 0\,,   
\end{align}
where we have defined:
\begin{equation}
    \mathcal{D}={F_{R\mathcal{G}}} G+{F_{RR}} R+1\,.\label{eq:T10}
\end{equation}

The next step it to evaluate the Eqs.(\ref{eq: hdot})\,-\,(\ref{eq: h4dot})  at redshift zero, and to use them in the previous equations (also evaluated at redshift zero) in order to find a relation between the derivatives of the $F(R,\mathcal{G})$ model and the cosmographic parameters. Thus, Eqs. \eqref{eq:T0}-\eqref{eq:T10},   evaluated at $z=0$, are:
\begin{align}
\frac{{F_{00}}+6 {H_0}^2 \left(24 {F_{R\mathcal{G}}} {H_0}^4 (2 j_0+q_0 (q_0+2)-2)+3 {F_{RR}} {H_0}^2 (2 j_0-q_0 (q_0+2)-3)-k {\Omega_m}+1\right)}{6 \mathcal{K}} =0\,,
\end{align}
\begin{align}
&\frac{{H_0}^2}{2\mathcal{K}} \biggl[48 {F_{R\mathcal{G}}} {H_0}^4 \left(-j_0 (4 q_0+5)-(q_0+6) q_0^2+q_0+s_0+6\right)+6 {F_{RR}} {H_0}^2 (-j_0+3 q_0 (q_0+3)+s_0+6)+3 k {\Omega_m}-\nonumber\\
&2 q_0-2\biggr]=0\,,
\end{align}
\begin{align}
&\frac{{H_0}^3}{2 \mathcal{K}^2} \biggl[-1152 {F_{R\mathcal{G}}}^2 {H_0}^8 \biggl(5 j_0^2+j_0 \left(2 q_0^3-11 q_0^2-5 q_0-s_0-6\right)+q_0 \left(-l_0+2 q_0^4-4 q_0^2+4 q_0 s_0+26 q_0+5 s_0+18\right)\biggr)-\nonumber\\
&6 {F_{RR}} {H_0}^2 \biggl(24 {F_{R\mathcal{G}}} {H_0}^4 \biggr(19 j_0^2-j_0 \left(12 q_0^3+56 q_0^2-13 q_0+3 s_0\right)-3 l_0 q_0+2 l_0-38 q_0^4-29 q_0^3+10 q_0^2 s_0+112 q_0^2+5 q_0 s_0+\nonumber\\
&12 q_0-12 s_0-48\biggr)-j_0 (3 k {\Omega_m}+5 q_0+3)+12 k q_0 {\Omega_m}-3 k {\Omega_m}+l_0-32 q_0^2-55 q_0-2 s_0-24\biggr)-24 {F_{R\mathcal{G}}} {H_0}^4 \biggl(8 j_0^2+\nonumber\\
&j_0 \left(-3 k {\Omega_m}+14 q_0^2+78 q_0+40\right)+2 \left(q_0^2 (22-3 k {\Omega_m})+l_0+17 q_0^3-2 q_0 (3 s_0+20)-4 (2 s_0+9)\right)\biggr)-\nonumber\\
&36 {F_{RR}}^2 {H_0}^4 \left(j_0^2+j_0 \left(2 q_0^2-8 q_0-s_0-15\right)-l_0 q_0+l_0+3 \left(9 q_0^3+14 q_0^2+q_0 (s_0-1)-4\right)\right)-2 j_0+9 k {\Omega_m}-6 q_0-4\biggr] =0\,,
\end{align}{}
where we have defined:
\begin{equation}
\mathcal{K} = 24 {F_{R\mathcal{G}}} {H_0}^4 q_0+6 {F_{RR}} {H_0}^2 (q_0-1)-1\,.   
\end{equation}
The final step is to solve the previous equations  with respect to the  derivatives of  $F(R,\mathcal{G})$ to get:
\begin{equation}
\frac{F(R_0,\mathcal{G}_0)}{H_0^2 } =  \frac{{\cal{P}}_1(q_0,
j_0, s_0, l_0) + {\cal{P}}_2(q_0, j_0, s_0, l_0, \Omega_M)}{{\cal{R}}(q_0, j_0, s_0,
l_0)} \ ,
\end{equation}
\begin{equation}
\frac{F_{RR}(R_0,\mathcal{G}_0)}{\left ( 6 H_0^2 \right )^{-1}} =  \frac{{\cal{P}}_3(q_0,
j_0, s_0, l_0) + {\cal{P}}_4(q_0, j_0, s_0, l_0, \Omega_M)}{{\cal{R}}(q_0, j_0, s_0,
l_0)} \ ,
\end{equation}

\begin{equation}
\frac{F_{R\mathcal{G}\mathcal{G}}(R_0,\mathcal{G}_0)}{\left ( 48 H_0^{4} \right )^{-1}} =  \frac{{\cal{P}}_5(q_0,
j_0, s_0, l_0) + {\cal{P}}_6(q_0, j_0, s_0, l_0, \Omega_M)}{{\cal{R}}(q_0, j_0, s_0,
l_0)} \ ,
\end{equation}
where we have defined\,:

\begin{align}
\label{eq:pol1}
 {\cal{P}}_1(q_0, j_0, s_0, l_0)  = & - 3 \biggl(8 j_0^3-24 j_0^2 q_0^3-94 j_0^2 q_0^2-64 j_0^2 q_0-8 j_0^2 s_0-14 j_0^2-8 j_0 l_0(q_0+1)+12 j_0 q_0^5-4 j_0 q_0^4+16 j_0 q_0^3+\nonumber\\
& 14 j_0 q_0^2 s_0+134 j_0 q_0^2-28 j_0 q_0 s_0+96 j_0 q_0-22 j_0 s_0+6 j_0+2 l_0q_0^3-6 l_0q_0^2-6 l_0q_0+2 l_0+24 q_0^6+6 q_0^5-\nonumber\\
&12 q_0^4 s_0-72 q_0^4-34 q_0^3 s_0-12 q_0^3+4 q_0^2 s_0+144 q_0^2+12 q_0 s_0^2+98 q_0 s_0+150 q_0+12 s_0^2+60 s_0+48\biggr)\,,
\end{align}

\begin{align}
{\cal{P}}_2(q_0, j_0, s_0, l_0, \Omega_M)  =  &
 3 \Omega_m k \biggl(16 j_0^3 +56 j_0^2  q_0+8 j_0^2  s_0-12 j_0^2 +8 j_0  l_0(q_0+1)+11 j_0  q_0^4+28 j_0  q_0^3+14 j_0  q_0^2 s_0+63 j_0  q_0^2+\nonumber\\
&48 j_0  q_0 s_0-57 j_0  q_0+12 j_0  s_0-24 j_0 +2  l_0q_0^3+12  l_0q_0^2+4  l_0q_0-3  l_0+6  q_0^5-33  q_0^4+8  q_0^3 s_0-\nonumber\\
&162  q_0^3-44  q_0^2 s_0-276  q_0^2-12  q_0 s_0^2-114  q_0 s_0-138  q_0-12  s_0^2-45  s_0-18\,, \biggr)
\end{align}

\begin{align}
 {\mathcal{P}}_3(q_0, j_0, s_0, l_0)  =&  2 j_0^2-12 j_0 q_0^3-56 j_0 q_0^2-72 j_0 q_0-2 j_0 s_0-30 j_0-2 l_0q_0-2 l-24 q_0^4-30 q_0^3+12 q_0^2 s_0+60 q_0^2+\nonumber\\
& 22 q_0 s_0+114 q_0+12 s_0+48\,,
\end{align}

\begin{align}
& {\mathcal{P}}_4 (q_0, j_0, s_0, l_0, \Omega_m)  =  \Omega_m k\left(12 j_0^2 +21 j_0  q_0^2+81 j_0  q_0+12 j_0 +3  l_0+36  q_0^3+6  q_0^2-18  q_0 s_0-114  q_0-15  s_0-54 \right)\,,
\end{align}

\begin{align}
& {\mathcal{P}}_5 (q_0, j_0, s_0, l_0)  =  -24 - 6 j_0 - 2 j_0^2 + 2 l_0- 78 q_0 - 12 j_0 q_0 + 2 l_0q_0 - 84 q_0^2 - 4 j_0 q_0^2 - 
 30 q_0^3 + 2 j_0 s_0 + 2 q_0 s_0\,,
\end{align}

\begin{align}
& {\mathcal{P}}_6 (q_0, j_0, s_0, l_0, \Omega_m)  =  \Omega_m k \left(15 j_0  q+30 j_0 -3  l_0+45  q_0^2+72  q_0-3  s_0+18 \right)
\end{align}

\begin{align}
{\mathcal{R}}(q_0, j_0, s_0, l_0)  = &  \biggl[4 j_0^3+j_0^2 \left(15 q_0^2+56 q_0-4 s_0+30\right)-j_0 \biggl(4 l_0(q_0+1)+16 q_0^4+32 q_0^3+q_0^2 (7 s_0+69)+6 q_0 (7 s_0+22)+\nonumber\\
& 24 (s_0+3)\biggr)+q_0^3 (-l_0+5 s_0+159)+q_0^2 (-9 l_0+46 s_0+306)+q_0 \left(6 \left(s_0^2+16 s_0+40\right)-8 l_0\right)-21 q_0^5+\nonumber\\
& 6 \left(s_0^2+8 s_0+12\right)\biggr]\,.
\label{eq:Rpol}
\end{align}

These are all the ingredient to construct our Gauss-Bonnet cosmography. See also \cite{2008PhRvD..78f3504C} for the case of $F(R)$ gravity.
\end{widetext}

\bibliographystyle{plain}

\end{document}